# The Impact of Medicaid Expansion on Medicare Quality Measures

Hala Algrain[1], Elizabeth Cardosa[1], Shekha Desai[1], Eugene Fong[1], Tanguy Ringoir[1], and Huthaifa I. Ashqar[1,2,*]

[1] Department of Professional Studies, University of Maryland, Baltimore County, MD, USA

[2] Arab American University, Jenin, Palestine

*Corresponding author (hiashqar@vt.edu)

## Abstract

The Affordable Care Act was signed into law in 2010, expanding Medicaid and improving access to care for millions of low-income Americans. Fewer uninsured individuals reduced the cost of uncompensated care, consequently improving the financial health of hospitals. We hypothesize that this amelioration in hospital finances resulted in a marked improvement of quality measures in states that chose to expand Medicaid. To our knowledge, the impact of Medicaid expansion on the Medicare population has not been investigated. Using a difference-in-difference analysis, we compare readmission rates for four measures from the Hospital Readmission Reduction Program: acute myocardial infarction, pneumonia, heart failure, and coronary artery bypass graft surgery. Our analysis provides evidence that between 2013 and 2021 expansion states improved hospital quality relative to non-expansion states as it relates to acute myocardial infarction readmissions (p = 0.015) and coronary artery bypass graft surgery readmissions (p = 0.039). Our analysis provides some evidence that expanding Medicaid improved hospital quality, as measured by a reduction in readmission rates. Using visualizations, we provide some evidence that hospital quality improved for the other two measures as well. We believe that a refinement of our estimation method and an improved dataset will increase our chances of finding significant results for these two other measures.

*Keywords*: ACA, Medicaid expansion, Hospital Readmission Reduction Program, policy analysis

# Introduction

The Affordable Care Act (ACA) provided states with federal financial support to expand Medicaid eligibility requirements beginning in January 2014 which resulted in 14 million previously uninsured Americans receiving coverage in states that chose to expand by the end 2021 (Finegold et al., 2021). The upsurge in coverage is associated with a subsequent improvement in both low income Americans' access to healthcare services and the quality of care they received (Mazurenko et al., 2018). Studies show a difference in hospital quality and clinical outcomes between expansion and non-expansion states for a variety of medical conditions and quality metrics (Crocker et al, 2019; Eguia et al., 2020; Hefei, 2019; Pickens et al, 2018). A systematic review of 404 Medicaid expansion studies found that insurance coverage, provider capacity, and health outcomes all improved in expansion states between 2014 and 2020 due to the policy change; the increase in insurance coverage for low income individuals in turn improved their "access to care, utilization of services, the affordability of care, and financial security" (Guth et al., 2020).

Medicaid expansion is also associated with improved financial performance and a decreased likelihood of hospital closures (Lindrooth et al., 2018). Hospitals in expansion states experienced a reduction in losses from uncompensated costs of care and a boost in Medicaid revenue and overall profit margins (Blavin, 2016). Improved hospital finances in turn promote higher safety and quality standards within the institution (Akinleye et al., 2019; Nguyen et al., 2016).

The Hospital Readmission Reduction Program (HRRP) is another ACA initiative aiming to improve quality while reducing costs by penalizing hospitals for excess risk-standardized 30-day readmission rates in the Medicare population admitted for six targeted conditions. Studies show the financial incentives stemming from the HRRP has been successful in reducing readmission rates and subsequently reimbursements costs associated with the target conditions (JAMA;).

In this study we examined: (a) the difference in uncompensated care in hospitals enrolled in the HRRP in states that expanded Medicare compared to those who did not, and (b) the difference in the excess readmission rates from the HRRP for acute myocardial infarction, heart failure, pneumonia, and chronic obstructive pulmonary disease between expansion and non-expansion states.

## Study Data and Methods

### Study population (overview)

For our analysis, the treatment group contains the 25 states that decided to expand Medicaid at the start of FY2014, and the control group contains the 12 states that haven't expanded Medicaid to this day. The remaining 13 states which chose to expand later were excluded (see Appendix for full lists).

### Study Variables

Outcomes (Difference in Changes in ERR)

The six primary outcomes were the ERR measures considered by HRRP. Beginning in fiscal year (FY) 2013 with the Medicaid expansion these included: acute myocardial infarction (AMI), heart failure (HF), and pneumonia (1). In FY 2015, chronic obstructive pulmonary disease (COPD) and elective primary total hip and/or total knee arthroplasty (THA/TKA) were added; in FY 2017, coronary artery bypass graft (CABG) surgery was added (1). According to the CMS, these six measures were specifically selected because of the potential breadth of their impact on the majority of Medicare beneficiaries and the variation in performance nationwide. This gives us an opportunity to improve the quality of care and save taxpayer dollars by incentivizing providers to reduce excess readmissions (2).

**Data Sources**

ERRs are calculated using the predicted to expected readmission rate. Predicted readmission rate is "the predicted 30-day readmission rate for a hospital based on its performance for its specific case mix (that is, the hospital-specific effect on readmission, provided in its discharge-level data in its Hospital-Specific Report)" (1). While the expected readmission rate is "the expected 30-day readmission rate for a hospital and is based on readmission rates at an average hospital with a patient case mix similar to that hopital's (that is, if patients with the same characteristics had been treated at an average hospital, rather than at that hospital)" (1).

IPUMS' National Historical Geographic Information System (NHGIS) was used to obtain geographic and demographic data. IPUMS provided access to the US Census data for 2013 at the census tract level. Census tract was chosen over larger geographies, such as ZIP code because each tract is limited to 8,000 people before it splits into two tracts (3). Below 1,200 people, tracts are recombined (3). Demographic data collected included: racial composition, per capita income, unemployment percentage, and the percentage of the population with a bachelor's degree or higher.

CMS Hospital CCN Lookup provided the name and location for each hospital and was accessed through the 2019 Medicare Inpatient Hospitals Access API (4). Data was also retrieved for states which participated in the Medicare expansion in 2014 (5) and a table with state names to state abbreviations was used to help merge this data with the ERR and demographic data (6). Obtaining the census tract of each hospital was required to merge the ERR and demographic data. The US Census Geocoding API was used to programmatically search for census tracts given a hospital's address (3). OpenStreetMap data was obtained through the Nominatim API to look up a hospital's latitude and longitude coordinates when the hospital address was not found directly. Then the Census Geocoder API was queried with hospital coordinates to find the census tract (7).

**Statistical Analysis**

The research method used was difference-in-difference analysis. This type of analysis has often been used in the literature to try to determine the effects of differing health policies across state lines (Hefei et al., 2019; Resnick, 2018; Sommers et al., 2017). The idea is to look at a health metric in a control group in which the health policy is not enacted, and to compare that to a health metric in a treatment group in which it is. We then assume that the health metric in the treatment group would have changed in the same way that it changed in the control group if the health policy was not enacted. This parallel trends assumption is used to account for the fact that the metric under consideration in the treatment group might have changed anyway, even if the health policy was not enacted. Using DiD estimation, we can evaluate the effect of Medicaid expansion on these hospital quality metrics.

We conducted a difference-in-difference (DiD) analysis to evaluate the effect of Medicaid expansion relative to non-expansion states during the same time period of 2013 - 2021. Implementation of this technique requires three elements:

- A dummy variable indicating whether the state in which the hospital is located is part of the treatment group or control group, with 1 indicating treatment and 0 indicating control.

- A variable that calculates the difference between ERRs for a metric between *post* and *pre*. The term *post* indicates the latest year for which we have data, which is 2021. The term *pre* indicates the base year for which we have data. This year is 2013 at the earliest, although data collection for some metrics began later.

- A variable that calculates the product of the previous two variables. This variable is called the interaction term or the DiD-estimator. This is our main variable of interest.

## Analysis and Results

We illustrate the differentials in ERRs over time for three metrics for the various states in the figures below, with the expansion states in blue and the non-expansion states in red. For each state and for each measure, the differentials are calculated by subtracting the ERR of 2013 from the ERR of 2021. Positive numbers therefore indicate that ERRs for that state increased and that quality therefore deteriorated in that state relative to CMS expectations. In the three figures below, we notice that the red non-expansion states tend to have positive differentials associated with them, and they tend to be more on the right than the left.

These visualizations provide some evidence that expansion states outperformed non-expansion states in hospital quality, but we know from the literature that there are factors that affect hospital quality that aren't picked up by these visualizations, most notably the demographic characteristics of the population in the tract in which the hospital is located, such as the racial composition, education level, and unemployment level. These factors can be controlled for using regression analysis, which we discuss next.

## Analysis and Results - Regressions

Despite the graphical evidence that quality improved more in expansion states than in non-expansion states for pneumonia and heart failure, the differential proved too small to show statistical significance in regressions. Those two regressions can be found in Table A1 and Table A2 in the Appendix. We did find statistically significant differences in quality between hospitals in expansion states and hospitals in non-expansion states over time for acute myocardial infarction, as evidenced by the negative and statistically significant coefficient on the interaction term in Table 1. A negative coefficient signifies fewer readmissions in expansion states, as hypothesized. There is no evidence that these results were impacted by COVID-19, since we re-ran the regression with 2019 as the *post* year compared to 2021, and no major differences in results were found. That regression can be found in Table A3 in the Appendix.

The average individual income by tract could not explain additional variation in our dependent variable over and above what was provided by our unemployment and education variables, and was therefore omitted from the regression. The racial composition variables were all positive and statistically significant. This provides additional evidence that hospitals in areas with a high proportion of minorities tend to provide worse quality care, exacerbating racial disparities observed in both morbidity and mortality. Somewhat surprisingly the coefficient on *asian* is positive - signifying worse hospital quality in areas with a high proportion of Asian Americans compared to our left-out category of mostly white individuals - even though, on average, Asian Americans tend to outperform all racial groups on measures of both socio-economic status and health.

Not all hospitals are equipped to perform CABG, resulting in far fewer observations to work with for regressions run on this variable. Data on CABG readmission rates started to be collected in 2017, which makes this our base year compared to 2013 for other regressions. Regression output can be found in Table 3.

**Table 2: Regression with excess readmission ratio for CABG as dependent variable.**

```
                        OLS Regression Results
==============================================================================
Dep. Variable:      err_for_cabg_2021   R-squared:                       0.556
Model:                            OLS   Adj. R-squared:                  0.552
Method:                 Least Squares   F-statistic:                     138.0
Date:                Mon, 13 Dec 2021   Prob (F-statistic):          1.13e-149
Time:                        16:29:56   Log-Likelihood:                 1031.1
No. Observations:                 889   AIC:                            -2044.
Df Residuals:                     880   BIC:                            -2001.
Df Model:                           8
Covariance Type:            nonrobust
==============================================================================
                     coef    std err          t      P>|t|      [0.025      0.975]
------------------------------------------------------------------------------------
const              1.0025      0.004    225.831      0.000       0.994       1.011
expanded_2014      0.0072      0.005      1.327      0.185      -0.003       0.018
err_diff_cabg      0.7275      0.029     25.025      0.000       0.670       0.785
err_interact_cabg -0.0869      0.042     -2.070      0.039      -0.169      -0.005
black              0.0050      0.005      1.103      0.270      -0.004       0.014
asian             -0.0423      0.020     -2.067      0.039      -0.083      -0.002
hispanic           0.0071      0.004      1.840      0.066      -0.000       0.015
unemployment      -0.0068      0.018     -0.370      0.712      -0.043       0.029
bachelor_or_higher -0.0091     0.008     -1.105      0.270      -0.025       0.007
```

We again find a statistically significant coefficient on our interaction term, indicating higher improvement in quality in expansion states compared to non-expansion states. Noteworthy is also the jump in adjusted R-squared, signifying that our variables do a much better job at explaining the variation in our dependent variable compared to the AMI regression. The significance of our unemployment and education variables suffered, likely due to the drop in observations. The variables were kept in the regression due to their theoretical basis and for consistency across analyses. Likewise, two of the three racial composition variables failed individual t-tests, but an F-test on the three racial composition variables illustrated that these variables belong in the regression. The statistically negative coefficient on *asian* is noteworthy despite being what we hypothesized beforehand, since this is different from what we found in our first regression.

## Discussion

### Limitations

There are some notable limitations of our dataset. The official ERR measures do not include Maryland because they have a separate quality measures program (Centers of Medicaid and Medicare, n.d. b). Additionally, any hospitals with fewer than 25 eligible discharges for a particular ERR measure will not be eligible for payment reduction for that measure (Centers of Medicaid and Medicare, n.d. b). Which also means that not all hospitals can perform every one of the ERR procedures and therefore, not all of them are included in each ERR comparison. States which chose to expand later after FY 2013 are not included in our analysis; we focused on the 25 that expanded in FY 2013 with only the ones that never expanded at all.

There were also some limitations in the final subset of hospital CCN's included. The final dataset uses hospitals that existed each year from 2013 to 2021, so any hospitals that either opened or closed during that time were excluded from the final dataset. Also, when merging the demographic and ERR data there were certain hospital CCN's that were dropped due to pre-processing challenges. The hospital CCN lookup was unable to find hospital names and addresses for 54 hospitals. The search for census tracts based on hospital addresses was unable to find 204 hospitals. From the US Census data there was completely missing demographic data for 225 census tracts.. These losses in data resulted from the 2,961 possible CCN's available from the CMS website that were in operation each year between 2013 and 2021, 2,578 hospitals were represented in our final dataset after the merging process.

We initially ran our analyses from 2013 to 2019 to exclude COVID-19, assuming it would skew the data. However, rerunning our analyses up to the year 2021 and comparing with results up to 2019 showed nearly the same results. No evidence that COVID-19 impacted results.

## Conclusion

The original three ERR measures we had decided to look at from the start of the Medicaid expansion in 2013 were: acute myocardial infarction, heart failure, and pneumonia. Our initial results seemed to have a significant reduction in readmissions for acute myocardial infarction but not the other two measures: heart failure and pneumonia. This begged the question of whether the observable difference here is based on the natural progression of a condition, such as, acute vs. chronic conditions? Acute myocardial infarction in layman's terms is a heart attack. Certainly that falls under "acute" compared to heart failure or pneumonia which can be long-term chronic conditions that the patient may not necessarily need to be admitted immediately to a hospital to treat, as opposed to a heart attack. There is some medical debate about what constitutes "chronic." Definitions range from "not cured once acquired or lasts ≥ 3 months" to a minimum duration of lasting "12 months or longer" (Goodman et al., 2013). Another one of the ERRs that may be considered "acute" is coronary artery bypass graf surgery. "The greatest risk is correlated with the urgency of operation, advanced age, and 1 or more prior coronary bypass surgeries" (Eagle et al., 1999). The "urgency of operation" depends on when it's discovered in the patient and level of blockage in the heart. For those patients who have a severe case needing immediate CABG, they are in a similar situation as AMI patients: acute case and inpatient hospital care. We decided to add CABG to our regression analysis and reran it and discovered a strong significance there as well, along with a marked reduction in readmissions. We hypothesize that acute conditions which necessitate an inpatient stay are more likely to benefit from the increased resources of Medicaid expansion, as opposed to chronic conditions which are more often outpatient and rely upon patient adherence to a doctor prescribed regimen. "Acute MI still carries a high mortality rate, with most deaths occurring prior to arrival to the hospital… Good outcomes are seen in patients who undergo early perfusion-thrombolytic therapy within 30 minutes of arrival or PCI within 90 minutes" (Mechanic et al., 2021). On the other hand, CABG has a higher preventable adverse event rate than other surgical procedures, requires highly specialized equipment, and an extensively trained team (Bachar & Manna, 2021). The American Heart Association (AHA) found that "poor teamwork and poor nontechnical skills have been shown to adversely affect patient outcomes" and suggested more team training and new simulation type training and assessments can improve outcomes (Wahr et al., 2013). This additional information may bolster our hypothesis that these acute type ERRs, which are inpatient procedures, may see the most improvement from Medicaid's expansion in the form of better equipment, training, and more experienced staff while under direct hospital supervision.

Our research hypothesis proposes that the financial improvements that resulted from medicaid expansion would also reflect on quality measures for the medicare population. We then saw significance in AMI readmission rates. AMI treatment is critically time-sensitive, this requires training and equipment that is costly which would logically give financially better off hospitals an advantage. This is similar for CABG best practices

Our results highlight an important criticism of the readmission program that we found in the literature. Chronic conditions like heart failure and chronic pneumonia have a natural history of slow progression that are more impacted by factors outside the hospital's control, such as, outpatient treatment and patient adherence. Contrast that with acute conditions like AMI and CABG, whose outcomes depend more on the hospital policies and quality improvement interventions.

In conclusion, our research contributes to the literature by providing evidence that the ACA, through its medicaid expansion policy, improved hospital quality more in expansion states than in non-expansion states. and not just for the Medicaid population, as previously shown in many other studies, but also for quality measures in the Medicare population.

**Future Directions**

We plan to run a more rigorous panel data analysis that would allow us to perform a regression on data collected at different points in time. Specifically, this would increase the size of our dataset by including the states that didn't expand on January 1, 2014, but expanded at a later point in time. With a larger dataset, a smaller differential in ERRs can be deemed statistically significant through increased statistical power. Our literature review showed that characteristics such as hospital size, tax exemption status, and financial performance were significant factors in how hospitals performed relative to quality programs - including the HRRP. Hospital-level data is available through the American Hospital Association. We have applied for a research grant from UMBC's Graduate Student Association to acquire this data. We believe that the larger panel dataset, together with the improvement of our model through the inclusion of additional relevant variables, will help us produce a paper ready for peer review.

# Appendix

25 early adopter states:

> Washington, Oregon, Nevada, California, Arizona, New Mexico, Colorado, North Dakota, Minnesota, Iowa, Illinois, Arkansas, Kentucky, West Virginia, New York, New Hampshire, Ohio, Connecticut, New Jersey, Delaware, Maryland, District of Columbia, Rhode Island, Massachusetts, Michigan

12 states without Medicaid expansion:

> Texas, Wyoming, South Dakota, Wisconsin, Kansas, Mississippi, Alabama, Georgia, South Carolina, North Carolina, Florida, Tennessee

**Table A1: Regression with ERR for pneumonia as dependent variable.**

```
                          OLS Regression Results
==============================================================================
Dep. Variable:     err_for_pneumonia_2021   R-squared:                   0.279
Model:                              OLS   Adj. R-squared:              0.277
Method:                   Least Squares   F-statistic:                 117.4
Date:                  Fri, 17 Dec 2021   Prob (F-statistic):       3.05e-166
Time:                          23:41:29   Log-Likelihood:             3342.6
No. Observations:                  2430   AIC:                         -6667.
Df Residuals:                      2421   BIC:                         -6615.
Df Model:                             8
Covariance Type:              nonrobust
==============================================================================
                          coef    std err          t      P>|t|      [0.025      0.975]
------------------------------------------------------------------------------
const                   0.9965      0.002    488.104      0.000       0.993       1.001
expanded_2014           0.0131      0.003      5.018      0.000       0.008       0.018
err_diff_pneumonia      0.3782      0.020     19.388      0.000       0.340       0.416
err_interact_pneumonia  0.0439      0.027      1.605      0.109      -0.010       0.098
black                   0.0115      0.002      6.392      0.000       0.008       0.015
asian                   0.0355      0.010      3.606      0.000       0.016       0.055
hispanic                0.0041      0.001      3.541      0.000       0.002       0.006
unemployment           -0.0201      0.006     -3.397      0.001      -0.032      -0.008
bachelor_or_higher     -0.0113      0.004     -2.630      0.009      -0.020      -0.003
==============================================================================
```

**Table A2: Regression with ERR for HF as dependent variable.**

```
                          OLS Regression Results
==============================================================================
Dep. Variable:          err_for_hf_2021   R-squared:                   0.293
Model:                              OLS   Adj. R-squared:              0.290
Method:                   Least Squares   F-statistic:                 125.3
Date:                  Fri, 17 Dec 2021   Prob (F-statistic):       4.79e-176
Time:                          23:41:29   Log-Likelihood:             3268.7
No. Observations:                  2431   AIC:                         -6519.
Df Residuals:                      2422   BIC:                         -6467.
Df Model:                             8
Covariance Type:              nonrobust
==============================================================================
                     coef    std err          t      P>|t|      [0.025      0.975]
------------------------------------------------------------------------------
const              0.9986      0.002    473.925      0.000       0.994       1.003
expanded_2014      0.0126      0.003      4.650      0.000       0.007       0.018
err_diff_hf        0.4406      0.021     21.417      0.000       0.400       0.481
err_interact_hf    0.0163      0.029      0.554      0.579      -0.041       0.074
black              0.0122      0.002      6.612      0.000       0.009       0.016
asian              0.0278      0.010      2.753      0.006       0.008       0.048
hispanic           0.0063      0.001      5.294      0.000       0.004       0.009
unemployment      -0.0180      0.006     -2.950      0.003      -0.030      -0.006
bachelor_or_higher -0.0188      0.004     -4.270      0.000      -0.027      -0.010
==============================================================================
```

**Table A3: Regression with ERR for AMI as dependent variable, with 2019 as *post*.**

```
                          OLS Regression Results
================================================================================
Dep. Variable:        err_for_ami_2019   R-squared:                     0.181
Model:                             OLS   Adj. R-squared:                0.178
Method:                  Least Squares   F-statistic:                   51.50
Date:                 Fri, 17 Dec 2021   Prob (F-statistic):         1.20e-75
Time:                         23:28:45   Log-Likelihood:               2731.4
No. Observations:                 1872   AIC:                          -5445.
Df Residuals:                     1863   BIC:                          -5395.
Df Model:                            8
Covariance Type:             nonrobust
================================================================================
                         coef    std err          t      P>|t|      [0.025      0.975]
--------------------------------------------------------------------------------
const                  1.0011      0.002    445.241      0.000       0.997       1.005
expanded_2014          0.0082      0.003      2.973      0.003       0.003       0.014
err_diff_2019_ami      0.3076      0.021     15.003      0.000       0.267       0.348
err_interact_2019_ami -0.0688      0.028     -2.416      0.016      -0.125      -0.013
black                  0.0116      0.002      5.390      0.000       0.007       0.016
asian                  0.0391      0.011      3.694      0.000       0.018       0.060
hispanic               0.0039      0.002      1.872      0.061      -0.000       0.008
unemployment          -0.0222      0.006     -3.460      0.001      -0.035      -0.010
bachelor_or_higher    -0.0163      0.005     -3.572      0.000      -0.025      -0.007
```